\documentclass[10pt,aps,prd,noshowpacs,eqsecnum,nofootinbib,twocolumn]{revtex4-1}
\usepackage{amsmath}
\usepackage{amsthm,amssymb}
\usepackage{mathrsfs,bbm}

\usepackage{nicefrac} 

\usepackage{longtable}

\usepackage{latexsym}

\usepackage{array}

\usepackage{color}

\begin{document}


\title{Seeking Texture Zeros in the Quark Mass Matrix Sector of the Standard Model}



\author{Yithsbey Giraldo}
\affiliation{Departamento de F\'\i sica, Universidad de Nari\~no, A.A. 1175, San Juan de Pasto, Colombia}
\date{\today}

\begin{abstract}
Here we show that the Weak Basis Transformation  is an appropriate mathematical tool that can be used to find texture zeros in the quark mass matrix sector of the Standard Model. So, starting with the most general quark mass matrices and taking physical data into consideration, is possible to obtain more than three texture zeros by any weak basis transformation. Where the most general quark mass matrices considered in the model,  were obtained through a special weak basis wherein the mass matrix $M_u$~(or $M_d$) has been taken to be diagonal and only the matrix $M_d$~(or $M_u$) is considered to be most general. 
\end{abstract}


\maketitle


\section{Introduction}
In the Standard Model~(SM), the masses of all quarks arise from the Yukawa Lagrangian
\begin{equation*}
 -{\cal L}_M=\bar u_RM_u u_L+\bar d_RM_dd_L+h.c.,
\end{equation*}
where the flavour structure of Yukawa couplings is not constrained by gauge symmetry
and, as a result, the up and down quark mass matrices, $M_u$ and $M_d$,  are arbitrary $3\times 3$ complex matrices, thus containing a total of 36 free parameters. A first simplification, without losing generality, the quark mass matrices can be considered Hermitian because of  the unitary matrix component, coming from the polar decomposition, can be absorbed in the right-handed quark fields. This immediately brings down the number of free parameters to 18 . This number is to be compared to the ten physical parameters corresponding to the six quark masses and four physical parameters of the Cabibbo-Kobayashi-Maskawa~(CKM) matrix. The above redundancy is closely related to the fact that one has the freedom to make weak-basis (WB) transformations under which the quark
mass matrices change but the gauge currents remain diagonal and real. 

For Hermitian quark mass matrices, one can further perform a common unitary  transformation on both left- and right-handed quark fields, which keeps the mass matrices to be Hermitian and has no physical effect, namely, the physical observables are unchanged under the WB transformation~\cite{b2,b_1}

\begin{subequations}
\label{2.1}
 \begin{align}
  M_u&\longrightarrow M_u^\prime=U^\dag M_u \,U,\\
M_d&\longrightarrow M_d^\prime=U^\dag M_d \,U,
 \end{align}
\end{subequations}
where $U$ is an arbitrary unitary matrix. It implies that the number of quark mass matrices representing the same model is infinity. 

Two sets of quark mass matrices related by a WB transformation obviously have the same physical content~\cite{b2}. Conversely, two sets of quark mass matrices having the same physical content are related by a WB transformation~\cite{b1}. That is, two sets of quark mass matrices have the same physical content if and only if they are related by a WB transformation.

On the other hand, in the absence of any viable theory for flavor physics, one usually resorts to phenomenological models such as texture specific mass matrices. Texture specific mass matrices were introduced explicitly by Fritzsch~\cite{bb1,bb1a,bb1b,bb1c}. In particular, Fritzsch-like texture specific mass matrices seem to be very helpful in understanding the pattern of quark mixings and CP violation. For details we refer the readers to~\cite{b2,b1,bb1,bb1a,bb1b,bb1c,b2a,c_3, c_4,c_6, c_6a, b4,b4a,b_1,c1x}.

This short paper is organized as follows: in Sect.~\ref{sIII}, we show the completeness of WB transformation by showing its hability to obtain different quark mass matrix representations and discuss its physical implications. Then we present in Sect.~\ref{sIV} a  kind of general quark mass matrices can be used as the starting point in the analysis,  and finally our conclusions are presented in Sect.~\ref{sV}.


\section{Completeness of WB Transformations}
\label{sIII}
Let us show  that the  WB  transformation method is complete in the sense that it  generates all possible quark mass matrix representations.  Let us consider first the Hermitian quark mass matrices indicated by {\small$(M_u,M_d)$}, and diagonalize them as follows
{\small
\begin{equation}
\label{19}
 U_u^\dag M_u U_u=D_u\quad\textrm{and}\quad U_d^\dag M_dU_d=D_d,
\end{equation}}
where CKM mixing matrix is given by
\begin{equation}
\label{20}
 V=U_u^\dag U_d.
\end{equation}
On the other hand, any other mass matrices {\small$(M^\prime_u,M^\prime_d)$} representing the same physical phenomenon give
{\small
\begin{equation}
\label{21}
 U_u^{\prime\dag} M^\prime_u U^\prime_u=D_u\quad\textrm{and}\quad U_d^{\prime\dag} M^\prime_dU^\prime_d=D_d,
\end{equation}}
and
\begin{equation}
\label{22}
 V=U_u^{\prime\dag} U^\prime_d.
\end{equation}
Equating the expressions in~(\ref{20}) and~(\ref{22}) yields
{\small
\begin{equation}
\label{23}
 U_u^\dag U_d=U_u^{\prime\dag} U^\prime_d\Rightarrow U^\prime_uU_u^\dag=U_d^\prime U_d^\dag.
\end{equation}}
And equating  expressions~(\ref{19}) and~(\ref{21}), gives respectively
{\small
\begin{equation}
 U_u^{\prime\dag}M_u^\prime U_u^\prime=U_u^\dag M_u U_u\quad\textrm{and}\quad U_d^{\prime\dag}M_d^\prime U_d^\prime=U_d^\dag M_d U_d,
\end{equation}}
where we find that the mass matrices $M_u$ and $M_d$ can be expressed in terms of the mass matrices $M_u^\prime$ and $M_d^\prime$ as follows
{\small
\begin{align}
\label{A7}
 M_u&=U_uU_u^{\prime\dag}M_u^\prime U_u^\prime U_u^\dag,\\
\label{26}
M_d&=U_dU_d^{\prime\dag}M_d^\prime U_d^\prime U_d^\dag.
\end{align}}
Using~(\ref{23}) into~(\ref{26}), we have
{\small
\begin{equation}
\label{A9}
 M_d=U_uU_u^{\prime\dag}M_d^\prime U_u^\prime U_u^\dag.
\end{equation}}
that together with~\eqref{A7} and given that $U=U_uU_u^{\prime\dag}$ is an unitary matrix allows us to state:

\noindent
\textit{  ``Two sets of quark mass matrices having the same 
 physical content are related by a WB transformation.''}

 So, starting from particular quark mass matrices, the WB transformation is able to find any other viable quark mass matrix configurations, i.e., if there exists a set of viable quarks mass matrices, it is certain that there is a unitary matrix leading to them. Although, the difficulty resides in finding an appropriated unitary matrix~\cite{b1}. 

Because some texture zeros must lie along the diagonal elements of both up and down Hermitian quark mass matrices, it implies that at least one, and at most two, of their eigenvalues must be negative~\cite{b2}. So, following the reasoning in the paragraph above,  this means that the relative sign of the quark mass parameters should be considered. Which implies a total of 36 independent initial quark mass matrices, depending what quark mass eigenvalues are negative. But, in the case of finding texture zeros, two negative eigenvalues can be reduced to only one by factoring a minus sign which can be absorbed into the quark mass  matrices, so that, for this case only 9 independent initial  quark mass matrices are considered, say each one with only a negative eigenvalue~\cite{b1}.

Hence, we are now able to explicitly construct texture  zeros in quark mass matrices through WB transformations. If these texture zeros exist, the WB transformation is able to find them.
Through WB transformations, Branco et al.~\cite{b2}  show that is always possible to find, at most, three zeros in quark mass matrices with no physical meanings. But, this does not restrict the number of zeros can be found by applying the WB transformation to mass matrices,  the case is that the model should be put into a physical context. Therefore, we have found  additional zeros~\cite{b1}~(four and even five texture zeros) by using the recent quark mass and mixing data. These additional zeros now have physical meanings because they were obtained from specific experimental data. 

With all this in mind, the question is, what quark mass matrices are to be used initially in order to find texture zeros through WB transformations. The answer is below.
\section{The Initial Quark Mass Matrices}
\label{sIV}
It is quite comfortable to use  as initial quark mass matrices the following structure

\begin{equation}
\label{3.1}
 M_u=D_u=\begin{pmatrix}
          \lambda_{1u}&0&0\\
0&\lambda_{2u}&0\\
0&0&\lambda_{3u}
         \end{pmatrix},\quad
%
 M_d=VD_dV^\dag,
\end{equation}
where the  up and down diagonal matrices  $D_u$ and $D_d$  contain the respective quark mass eigenvalues, and $V$ is the usual  quark  CKM mixing  matrix. The calculus of texture zeros is facilitated by using these initial quark mass matrices, because we have simultaneously available  the quark masses and the CKM matrix elements. The starting matrices~\eqref{3.1}, used in papers like~\cite{b2,b1,b3},  are as general as any other one. The reason is that starting from arbitrary Hermitian matrices $M_u$ and $M_d$, and using their respective diagonalizing matrices  $U_u$ and $U_d$,  and performing a  WB transformation~\eqref{2.1} using for this case  the unitary matrix $U=U_u$, we have
{\small
\begin{equation*}
 \begin{split}
  M_u\longrightarrow M'_u&=U^\dag_u\, M_u\, U_u=D_u,
\\
M_d=U_d\,D_d \,U_d^\dag\longrightarrow M'_d&=U^\dag_u\,(U_d \,D_d \,U_d^\dag)\,U_u,
\\
&=(U^\dag_u\,U_d)\,D_d\,(U_d^\dag \,U_u),
\\
&=V \,D_d \,V^\dag ,
 \end{split}
\end{equation*}}
where the CKM mixing matrix $V=U^\dag_u\, U_d$ was considered.  Additionally, note also that the three no physical texture zeroes mentioned above appear also in~\eqref{3.1}.

The other possibility that also works well is derived as follows
{\small
\begin{equation*}
 \begin{split}
  M_d\longrightarrow M'_d&=U^\dag_d\, M_d\, U_d=D_d,
\\
M_u=U_u\,D_u \,U_u^\dag\longrightarrow M'_u&=U^\dag_d\,(U_u \,D_u \,U_u^\dag)\,U_d,
\\
&=(U^\dag_d\,U_u)\,D_u\,(U_u^\dag \,U_d),
\\
&=V^\dag \,D_d \,V ,
 \end{split}
\end{equation*}}
such that we obtain a similar quark mass matrix  structure
\begin{equation}
\label{3.2}
 M_u=V^\dag D_uV,\quad
%
 M_d=D_d=\begin{pmatrix}
          \lambda_{1d}&0&0\\
0&\lambda_{2d}&0\\
0&0&\lambda_{3d}
         \end{pmatrix},
\end{equation}
which is as general as the previous one.

\section{Conclusions}
\label{sV}
To begin with, the WB transformation is complete, so we can find all possible quark mass matrices representing the model by starting from  specific quark mass matrix configurations. It is important to mention that, in the SM, is always possible to find a maximum of three no physical vanishing elements in the quark mass matrices by performing a WB transformation. In the process does not matter the value of physical quantities. But if we want  to find additional texture zeros is necessary to take into account physical considerations. 

Another important result, emphasized by other authors, is that the quark mass matrix  structure given in~\eqref{3.1}~(or \eqref{3.2}), which was called in my  paper~\cite{b1} as the{ \it u-diagonal representation}~(or the{ \it d-diagonal representation}), is so general as any other one. These matrices are deduced from a WB transformation and  have the advantage of having available the quark masses and the CKM matrix elements.

Taking into account the two paragraphs above, we have a solid way to find texture zeros in the quark mass matrix sector of the SM: it consists, initially, by choosing the general quark mass matrix structure~\eqref{3.1}~(or~\eqref{3.2}) and applying them appropriated  WB transformations, which implies that four and five texture zeros, if they exist, is going appear~\cite{b1}. 


\section*{Acknowledgments}
This work was partially supported by Department of Physics in the Universidad de Nari\~no, approval Agreement Number 009.




\nocite{*}
\bibliographystyle{unsrt}

\bibliography{yg}
\end{document}